\documentclass[12pt]{article}
\usepackage[margin=0.95 in]{geometry}
\usepackage{amsmath} 
\usepackage{amssymb,amsfonts}
\usepackage[all]{xy}
\usepackage{graphicx}

\usepackage{amsmath}
\usepackage{amssymb}
\usepackage{float}
\usepackage{array}
\usepackage{tikz}
\usepackage{mathtools}
\usepackage{mathrsfs} 
\usepackage[hidelinks]{hyperref}
\usepackage{cite}
\numberwithin{equation}{section}
\usepackage[utf8x]{inputenc}

\newcommand{\be}{\begin{equation}}
\newcommand{\ee}{\end{equation}}
\def\bea{\begin{eqnarray}}
\def\eea{\end{eqnarray}}
\newcommand{\qst}{q_{\text{s.t.} }}
\newcommand{\taust}{\tau_{\text{s.t.} }}

\newcommand{\ii}{\mathrm{i}}

\renewcommand{\b}{\beta}

\newcommand{\qbarst}{\bar{q}_{\text{s.t.} }}
\newcommand{\taubarst}{\bar{\tau}_{\text{s.t.} }}

\newcommand{\nubarst}{\bar{\nu} }

\numberwithin{equation}{section}
\numberwithin{table}{section}

\begin{document}

\hypersetup{pageanchor=false}
\begin{titlepage}
\vbox{
    \halign{#\hfil         \cr
           } % end of \halign
      }  % end of \vbox
\vspace*{15mm}
\begin{center}
{\Large \bf 
String Scale  Thermal Anti-de Sitter Spaces
}

\vspace*{15mm}

{\large Sujay K. Ashok$^a$ and Jan Troost$^b$ }
\vspace*{8mm}

$^a$The Institute of Mathematical Sciences, \\
          Homi Bhabha National Institute (HBNI),\\
		 IV Cross Road, C.I.T. Campus, \\
	 Taramani, Chennai, India 600113

\vspace{.8cm}

$^b$ Laboratoire de Physique de l'\'Ecole Normale Sup\'erieure \\ 
 \hskip -.05cm
 CNRS, ENS, Universit\'e PSL,  Sorbonne Universit\'e, \\
 Universit\'e de Paris 
 \hskip -.05cm F-75005 Paris, France	 
\vskip 0.8cm
	{\small
		E-mail:
		\texttt{sashok@imsc.res.in, jan.troost@ens.fr}
	}
\vspace*{0.8cm}
\end{center}

\begin{abstract}

We study finite temperature string scale $AdS_3$  backgrounds.
One background is $AdS_3 \times S^1 \times T^2$ in which the anti-de Sitter space-time and the circle are at the  radius $\sqrt{\alpha'}$. Using path integral techniques, we show that the  bulk spectrum includes a continuum of states as well as  Ramond-Ramond ground states  that agree with those of the symmetric orbifold of the two-torus after second quantization.
We also examine the one-loop free energy of  the background  $AdS_3 \times S^1$ at curvature radius $\sqrt{2 \alpha'/3}$. In the space-time NSNS sector, the string theory spontaneously breaks conformal symmetry as well as R-charge conjugation symmetry. We prove that the minimum in the boundary energy is reached for a singly wound string. In the RR sector, we classify the infinite set of ground states with fractional R-charges. Moreover, we remark on the behaviour of critical temperatures as the curvature scale becomes smaller than the string scale.
\vspace{1em} \\ 
In an appendix, we derive the Hawking-Page transition in string theory by integrating a world sheet one-point function.

\end{abstract}

\end{titlepage}
\hypersetup{pageanchor=true}

\tableofcontents

%
% To do:
%
%
%

\section{Non-Critical Anti-de Sitter Space-times}

Quantum theories of gravity may all be holographic
\cite{tHooft:1993dmi,Susskind:1994vu}. There is good evidence that this is true for string theories in anti-de Sitter space-times \cite{Maldacena:1997re}.
String theories in three-dimensional anti-de Sitter space-times with Neveu-Schwarz Neveu-Schwarz flux are exceptional in that they are to a considerable extent solvable to all orders in the curvature radius. In this paper, we wish to exploit the latter feature to the fullest by analyzing string theories that live in anti-de Sitter space-times that have string scale curvature, or that are even more highly curved. A very interesting example of this type is the theory  at the string scale 
with ${\cal N}=4$ superconformal symmetry in space-time as
described in the hybrid formalism \cite{Eberhardt:2019ywk}. The extended supersymmetry combined with the scale of curvature render the spectrum almost topological, and the theory becomes highly computable \cite{Eberhardt:2019ywk}.  Examples in the NSR formalism, in particular with ${\cal N}=4 $ superconformal symmetry, were studied in detail in \cite{Giribet:2018ada}.

It was argued in 
\cite{Giveon:2005mi} that at a radius of curvature that is smaller than the string scale, the $SL(2,\mathbb{C})$ invariant vacuum of the  theory becomes non-normalizable. There is therefore a  transition at the string scale radius, at which the effective central charge of the theory follows a different trajectory due to the spontaneous symmetry breaking and the non-normalizability of the global $AdS_3$ state.
Moreover, the thermodynamics of string theory in Neveu-Schwarz Neveu-Schwarz $AdS_3$ is such that there are three critical temperatures. There is a Hawking-Page transition between the thermal $AdS_3$ phase and a BTZ black hole \cite{Hawking:1982dh,Witten:1998zw,Maldacena:1998bw}.
In addition there is  a Hagedorn transition both for perturbative string theory in euclidean $AdS_3$ and for string theory in the BTZ black hole background
\cite{Berkooz:2007fe}. The  three critical temperatures associated to these  transitions coincide at the critical radius of curvature $R=\sqrt{\alpha'}$. 

Thus, we are motivated to study anti-de Sitter string theories with a curvature radius equal to the string scale or smaller still. When we have such a string theory, the high curvature of the $AdS_3$ space-time renders the theory non-critical (in the traditional sense of changing the number of macroscopic space-time dimensions -- the string theories remain anomaly free). One approach to building such string theories is to consider a flat space non-critical superstring theory \cite{Bilal:1986ia,Bilal:1986uh,Mizoguchi:2000kk,Murthy:2003es} and to add a density of fundamental strings to these backgrounds. Then taking the near string limit results in a highly curved anti-de Sitter geometry with Neveu-Schwarz Neveu-Schwarz flux \cite{Giveon:1999zm}.

Simple examples of such models are the backgrounds:
\begin{eqnarray}
(0)  &:& AdS_3^{(2)} \times S^1 \times T^4 \nonumber \\
(1) &:& AdS_3^{(1)} \times S^1 \times T^2 \nonumber \\
(2) &:& AdS_3^{(\frac{2}{3})} \times S^1 
%
%
%(3)
\, 
\end{eqnarray}
with respectively eight, six and four fermionic super partners to the  bosonic world sheet coordinates. We have indicated as an upper index the supersymmetric level $k$ of the Wess-Zumino-Witten model that describes the $AdS_3$ background with Neveu-Schwarz Neveu-Schwarz flux. We expect the total number of supercharges in these backgrounds in the near horizon limit to be sixteen, eight and eight respectively. In type IIB string theory, we expect them to be dual to a  ${\cal N}=(4,4)$ superconformal theory and  two ${\cal N}=(2,2)$ models in the near string limit.  The background labelled $(0)$ is above the string scale and can be analyzed by largely the same techniques as critical string backgrounds. 
The other two models we find particularly interesting and they are the subject of this paper. Background (1) is at the critical radius  and it is the subject of  section \ref{StringScale}. Model (2) is curled up more strongly than the string scale and we analyze its peculiar properties in section \ref{AdS3S1}. While both models have been mentioned in the literature, we study them in more detail and illuminate their spectra using path integral techniques. 
We will revisit slightly more general models in section \ref{BeyondStringScale} that  interpolate between backgrounds (1) and (2)  to a degree and that allow to discuss the finite temperature thermodynamics in a broader setting.
We draw conclusions in section \ref{Conclusions}.

Finally,  Appendix \ref{HPWS} provides a   world sheet derivation of the Hawking-Page transition from a one-point function in perturbative string theory. 

\section{An Exactly String Scale Anti-de Sitter Space}
\label{StringScale}
In this section, we study string theory on $AdS_3 \times S^1 \times T^2$ with six fermions. The $AdS_3$ space-time is described by a world sheet conformal field theory with an $sl(2,\mathbb{R})$ current algebra at  supersymmetric level $k=1$. The radius of curvature is equal to the string scale $\sqrt{\alpha'}$. The circle is at the self-dual radius $\sqrt{\alpha'}$. The $AdS_3$ space-time is at the correspondence point \cite{Horowitz:1996nw} at which the $SL(2,\mathbb{C})$ invariant ground state is at the cusp of normalizability \cite{Giveon:2005mi}.  At this critical radius,  the Hagedorn temperature of the euclidean $AdS_3$ space-time and of its dual BTZ euclidean black hole coincide with the Hawking-Page transition temperature  \cite{Berkooz:2007fe}.    The total central charge of the matter theory is 
\be
c=\left(3 + \frac{6}{k}\right)_{k=1} + 3  +6 \times \frac{1}{2} = 15~,
\ee
and the super string theory that includes the ghosts and super ghosts is conformal anomaly free. It is useful to think of the light cone directions as including the time-direction in $AdS_3$ as well as the $S^1$ direction. The transverse excitations then lie in  a level one $sl(2,\mathbb{R})/u(1)$ coset and the two-torus directions. There is  still a sum over zero modes in the light-cone directions. The background can be thought off as corresponding to a non-critical string background \cite{Bilal:1986ia,Bilal:1986uh,Mizoguchi:2000kk,Murthy:2003es}  supplemented with a density of fundamental strings, described in the limit where we are very near the fundamental strings \cite{Giveon:1999zm}. We study the partition function of the perturbative fluctuations around this background, after thermal compactification. The one-loop partition function is thus a sum over maps from a toroidal world sheet to a euclidean thermal $AdS_3$ space-time with torus boundary  \cite{Maldacena:2000kv}. From this one loop amplitude and its integrand we can read off the on-shell and off-shell spectrum of fluctuations. We will exhibit interesting properties of the spectrum.

 A model where the two-torus is replaced by four fermions and with a different amount of space-time supersymmetry has been studied  in \cite{Giribet:2018ada}. A model at the correspondence point described in a hybrid formalism and with ${\cal N}=(4,4)$ superconformal symmetry has been analyzed in depth in \cite{Eberhardt:2019ywk}. Indeed,  the detailed model we pick results from  a number of simple choices and we would like to stress that  a wealth of variations remains possible.

\subsection{The Single Particle Free Energy}

The one loop string amplitude is given by an integral over the fundamental domain $F_0$ of the complex structure moduli space of the torus \cite{Polchinski:1998rq}. 
The world sheet conformal field theory partition functions  contain the bosonic thermal $AdS_3$ degrees of freedom as well as decoupled bosons and fermions. Because of their decoupled nature, the latter degrees of freedom behave as in the flat space non-critical partition functions   \cite{Bilal:1986ia,Bilal:1986uh,Mizoguchi:2000kk,Murthy:2003es}. 
We thus propose the one-loop vacuum amplitude:
\begin{align}
\label{Zloopk1}
Z = \frac{1}{2\pi}\int_{F_0} \frac{d^2\tau}{\tau_2^2}~Z_{AdS_3}~ & Z_{T^2}~Z_{gh}  \nonumber \\
\frac{1}{4|\eta(\tau)|^6} 
\Big( 
& \left |\Theta_{1,1}(0, \tau) (\theta_3^2(0, \tau)+ \theta_4^2(0, \tau))
- \Theta_{0,1}(0, \tau) \theta_2^2(0, \tau) \right|^2 %\right. 
\nonumber \\
%\left.
+ & \left | \Theta_{0,1}(0, \tau) (\theta_3^2(0, \tau)- \theta_4^2(0, \tau))
- \Theta_{1,1}(0, \tau) \theta_2^2(0, \tau)\right|^2
\Big)~.
\end{align}
We made use of the definition of the theta-functions $\Theta_{m,l}$ at level $l$
\begin{equation}
\Theta_{m,l}(q,z) = \sum_{n \in \mathbb{Z}}
q^{
l(n+\frac{m}{2l})^2}
z^{l(n+\frac{m}{2l})}~,
\end{equation}
 with the notations $q=e^{2 \pi i \tau}$ and $z=e^{2 \pi i \nu}$ and we use standard conventions for the  functions $\theta_i$ \cite{Polchinski:1998rq}.
As in non-critical string backgrounds, the type II GSO projection not only acts on the world sheet fermions, but also by translation along the circle, thus intertwining the fate of the fermions with that of the circle zero modes. As a consequence of theta-function identities, one can check that the one loop string amplitude is modular invariant and vanishes, ensuring that the background is supersymmetric.

We wish to dress our partition functions with fugacities that will make it possible to both read off the space-time conformal dimensions  as well as the R-charges of the on-shell physical excitations \cite{Maldacena:2000kv,Ashok:2020dnc}. We turn on the complex fugacities $(\taust, \taubarst)$ associated to the left/right energy and angular momentum in space-time \cite{Maldacena:2000kv}:
\be
\label{boundarytau}
\taust = \frac{1}{2\pi}(\beta + \ii\,\beta\,\mu)~,
\ee
where $\beta$ is the inverse temperature and $\mu$ is a chemical potential for spin.
We also turn on real fugacities $(\nu, \bar \nu)$ that couple to the space-time left/right $U(1)_R$ generators \cite{Ashok:2020dnc}. At non-zero fugacity, an extra world sheet sector in the NSR formalism contributes and the theta-function identity that captures space-time supersymmetry also acquires a non-zero right hand side.  The latter  will represent the  degrees of freedom in a more manifestly supersymmetric Green-Schwarz form. Thus, we  make use of generalized abstruse theta-function identities:
\begin{small}
\begin{align}
&\Theta_{1,1}(\nu_3, \tau) (\theta_3(\nu_1, \tau)\theta_3(\nu_2, \tau)+ \theta_4(\nu_1, \tau)\theta_4(\nu_2, \tau))
-\Theta_{0,1}(\nu_3, \tau) (\theta_2(\nu_1, \tau) \theta_2(\nu_2, \tau) -\theta_1(\nu_1, \tau) \theta_1(\nu_2, \tau) ) \nonumber \\
& =2 \, \Theta_{0,1}(\nu_1-\nu_2,\tau)\theta_1\left(\frac{\nu_1+\nu_2+\nu_3}{2},\tau\right)\theta_1\left(\frac{\nu_1+\nu_2-\nu_3}{2},\tau\right)~.
\\
&\Theta_{0,1}(\nu_3, \tau) (\theta_3(\nu_1, \tau)\theta_3(\nu_2, \tau)- \theta_4(\nu_1, \tau)\theta_4(\nu_2, \tau))
-\Theta_{1,1}(\nu_3, \tau) (\theta_2(\nu_1, \tau) \theta_2(\nu_2, \tau) +\theta_1(\nu_1, \tau) \theta_1(\nu_2, \tau) ) \nonumber \\
&=-2 \, \Theta_{1,1}(\nu_1-\nu_2,\tau)\theta_1\left(\frac{\nu_1+\nu_2+\nu_3}{2},\tau\right)\theta_1\left(\frac{\nu_1+\nu_2-\nu_3}{2},\tau\right)~.
\end{align}
\end{small}
When the fugacities are all set to zero, the right hand side of each of the two identities vanishes  and the combinations that appear on the left hand side are precisely those that feature in the partition function  \eqref{Zloopk1} leading to the vanishing one-loop amplitude on thermal $AdS_3$ with periodic boundary conditions for the fermions along the thermal circle.
Considering the origin of the various factors in the partition function in the NSR formalism \cite{Maldacena:2000kv,Ashok:2020dnc} as well as the natural identification of the space-time R-charge with a multiple of the $S^1$ momentum \cite{Giveon:2003ku}, we are led to choose the fugacities as: 
$\nu_1=\taust$, $\nu_2=0$ and $\nu_3=2 \nu$.
The  twisted partition function in the space-time NSNS sector and in the manifestly supersymmetric formulation is then
\begin{multline}
Z(\taust, \nu) = \frac{1}{2\pi}\int_{F_0} \frac{d^2\tau}{\tau_2^2}~Z_{AdS_3}(\taust, \taubarst)~Z_{T^2}~Z_{gh}~\frac{1}{|\eta(\tau)|^4}\\
\bigg(\left |\frac{\Theta_{0,1}(\taust, \tau)}{\eta(\tau)}\right|^2 + \left|\frac{\Theta_{1,1}(\taust, \tau)}{\eta(\tau)}\right|^2   \bigg)\left| \theta_1\left(\frac{\taust}{2}+\nu,\tau\right)\theta_1\left(\frac{\taust}{2}-\nu,\tau\right)\right|^2~.
\end{multline}
The factor in the parentheses in the second line is a twisted boson partition function  at self-dual radius. We  proceed as in \cite{Maldacena:2000kv, Ashok:2020dnc}. We  use the unfolding trick to extract the single string free energy. We also spectral flow in the boundary theory by 
\be
\nu \rightarrow \nu - \frac{\taust}{2} \qquad \nubarst \rightarrow \nubarst - \frac{\taubarst}{2}~,
\ee
and, up to an overall factor, find the single string contribution to the free energy $f$ in the Ramond-Ramond sector:
\begin{multline}
f(\b,\mu, \nu) = \frac{1}{2\pi} \int_{-\frac12}^{\frac12} d\tau_1\int_0^{\infty} \frac{d\tau_2}{\tau_2^\frac{3}{2}}~\frac{e^{-\frac{\b^2}{4\pi\tau_2}}}{|\theta_1(\taust,\tau) |^2}Z_{T^2}\\
\bigg(\left |\frac{\Theta_{0,1}(\taust, \tau)}{\eta(\tau)}\right|^2 + \left|\frac{\Theta_{1,1}(\taust, \tau)}{\eta(\tau)}\right|^2   \bigg)
\left| \theta_1(\nu,\tau)\theta_1(\taust-\nu,\tau)\right|^2~.
\end{multline}
Following the logic in \cite{Maldacena:2000kv,Ashok:2020dnc} to which we refer the reader for details, it is possible to separate out the contribution of the discrete states to the single string free energy  while exhibiting the off-shell Hilbert space of the theory:
\begin{align}
f_{\text{disc}}(\b,\mu, \nu) =&    \int_{-\frac12}^{\frac12} d\tau_1\int_0^{\infty} \frac{d\tau_2}{\tau_2} \sum_{r,f_1,f_2,n_L,n_R,w}e^{\pi\ii(f_1+f_2-\bar f_1+\bar f_2)}\int_{(\frac12,1]}~\frac{dj}{\pi}~
q^{L_0-\frac{1}{2}}~ \bar q^{\bar L_0 - \frac12} \cr
&\hspace{0.1cm}
d(j,r,f_i,n_{L,R},q) \, 
\qst^{ j + r+f_1-\frac12+ \frac{n_L}{2}}~ \qbarst^{ j + \bar r+\bar f_1-\frac12+ \frac{n_R}{2}}~z^{w+f_2-f_1} \bar z^{w+\bar f_2-f_1}~.
\label{fdisc}
\end{align}
We indicated by $d$ the degeneracies of the states as a function of their  quantum numbers. They can be derived for the discrete representations and can be shown to reassemble into discrete characters \cite{Ashok:2020dnc}. 
The left-moving worldsheet Virasoro generator $L_0$ is 
\begin{align}
\label{wsVirasoro}
L_0 =& -j(j-1)  -w\big(j+r + f_1-\frac12 +\frac{n_L}{2}\big)+\sum_{i=1}^2 \frac12(f_i-\frac12)^2+\frac{n_L^2}{4}~,
\end{align} 
and similarly for the right-movers. 
The first terms arise from a spectral flowed bosonic $sl(2,\mathbb{R})$ model at level $k=1$, while the following terms correspond to four transverse fermions as well as a compact boson at self-dual radius.\footnote{The left and right moving momenta for a compact boson are defined as \cite{Polchinski:1998rq}
\be
\label{pLpRcircle}
p_L = \frac{\widetilde{n}}{R} + \frac{\widetilde{w}R}{\alpha'} \qquad p_R = \frac{\widetilde{n}}{R} - \frac{\widetilde{w}R}{\alpha'}~.
\ee
The self-dual radius $R=\sqrt{\alpha'}$ equals $\sqrt{2}$ when we set $\alpha'=2$. We define the integers $n_L = \widetilde{n}+ \widetilde{w}$ and $n_R=\widetilde{n} - \widetilde{w}$ so that $\sqrt{2}p_L = n_L$ and$\sqrt{2}p_R = n_R$. The parity of $n_{L,R}$ is correlated.} The fermion numbers $f_i$ count both the worldsheet and space-time spinors modulo two. Apart from the discrete contribution, the terms corresponding to the continuous representations of the $AdS_3$ isometry group can also be identified \cite{Maldacena:2000kv}.

\subsection{The Ramond-Ramond Ground States}

The left moving R-charge $Q^R_{\text{s.t.}}$ and conformal dimension $H$ of an excitation in  the boundary theory arising from the discrete representations are given respectively by the exponents of $\qst$ and $z$ in equation (\ref{fdisc}):
\begin{align}
H &= j + r+f_1-\frac12+ \frac{n_L}{2}~, \qquad Q^R_{\text{s.t.}} = w+f_2-f_1~.
\end{align}
Our next goal is to find the minimum value of the conformal dimension $H$ among the on-shell single string excitations in the space-time Ramond-Ramond sector. It is known that  the optimal value for the quantum number $r$ (which is a $sl(2,\mathbb{R})$ spin component relative to a discrete representation lowest weight state) is $r=0$ \cite{Ashok:2020dnc}.  We solve for the $sl(2,\mathbb{R})$ spin $j$ by using the on-shell condition $L_0=\frac12$, where the worldsheet conformal dimension $L_0$ is given in \eqref{wsVirasoro}. The space-time left-moving conformal dimension $H$ is then expressed in terms of the integer quantum numbers $(f_1, f_2, n_L)$. The extremum of the dimension $H$  occurs for fermion number $f_1=0$ and the integer value $n_L = -1$. For any spectral flow number $w$, we find that the minimum  $H=0$ is realized by two states with the quantum numbers  $f_2\in\{0,1\}$ and the $sl(2,\mathbb{R})$ spin $j$ at the boundary $j=1$ of the allowed range $1/2<j \le 1$. We table the Ramond ground states of the boundary theory in the left-moving sector along with their R-charges $Q^R_{\text{s.t.}}$:
\begin{table}[H]
    \centering
    \begin{tabular}{c|c}
       $f_2$  & $Q^R_{\text{s.t.}}$  \\
       \hline
       $ 0 $ & $w $\\
       $ 1 $ & $w+1$\\
    \end{tabular}
    \caption{The (boundary) R-charge of the Ramond ground states for a given value of the positive spectral flow number $w$ and fermion number $f_2$. }
    \label{tab:RRgroundstates:T2}
\end{table}
\noindent
The R-charges of the Ramond ground states come in pairs $(w,w+1)$ for $w \ge 0$. These are also twice the left-moving conformal dimensions of the corresponding operators in the NSNS sector. The right-moving sector has equal  winding number and an independent fermion number $\bar{f}_1$. The  single particle ground state spectrum at $w=0$ corresponds to the Hodge diamond of  a two-torus $T^2$. The second quantized  Ramond-Ramond ground states which allow for any number of excitations for all windings $w$ thus correspond to the symmetric product of the two-torus ground states.  Beyond the Ramond-Ramond ground states that we determined, there are further discrete states as well as a continuum. The bottom of the continuum gives rise to a tower of massless higher spin fields \cite{Gaberdiel:2017oqg}. 

We have  provided a twisted partition function for our model, from which we derived the complete Ramond-Ramond ground state spectrum. Our spectrum differs from the spectrum of other models in the literature \cite{Giribet:2018ada,Eberhardt:2019ywk}. We believe it provides a useful alternative to results obtained directly from the analysis of world sheet vertex operators \cite{Giveon:1999zm,Argurio:2000tb} and  provides a proof of the completeness of the classification. It would certainly be interesting to understand more properties of this model, including the three-point functions of chiral primaries, and their comparison to proposed dual conformal field theories. In particular, it will be interesting to understand to what extent the topological features of the ${\cal N}=4$ superconformal hybrid string at the correspondence point \cite{Eberhardt:2019ywk} continue to hold in this ${\cal N}=2$ superconformal model with a continuous part to its spectrum. Manifestly, it will be useful to exploit those techniques that apply to all models at supersymmetric level $k=1$.

\section{Strings in \texorpdfstring{$AdS_3 \times S^1$}{}}
\label{AdS3S1}
In this section, we study a background with a curvature scale smaller than the string scale. 
We present the thermal partition function on the background $AdS_3 \times S^1$ supplemented with four fermions. 
 The three-dimensional anti-de Sitter space has a  curvature radius  $R=\sqrt{2 \alpha'/3}$ equal to the radius of the circle. The supersymmetric level of the $sl(2,\mathbb{R})$ Wess-Zumino-Witten model is $k=2/3$. The total matter central charge is  again critical:
\be
c= \left(3 + \frac{6}{k}\right)_{k=\frac23} + 1  + 4\times \frac12 = 15.
\ee
  We calculate the one-loop single string free energy  and discuss the peculiar features of this background at curvature radius below the string scale.

\subsection{The One Loop Free Energy}
The one-loop vacuum amplitude in the NSR formalism is again modelled on the non-critical superstring partition function in flat space \cite{Bilal:1986ia,Bilal:1986uh,Mizoguchi:2000kk,Murthy:2003es}
\begin{equation}
\label{Zloopk23}
Z = \frac{1}{2\pi}\int_{F_0} \frac{d^2\tau}{\tau_2^2}~Z_{AdS_3}~Z_{gh}
\frac{1}{4|\eta(\tau)|^4} \left | \theta_{3}(0, \tau) \theta_3(0, \frac{\tau}{3})- \theta_4(0, \tau)\theta_4(0, \frac{\tau}{3})
- \theta_2(0, \tau)\theta_2(0,\frac{\tau}{3}) \right|^2~.
\end{equation}
The partition function is modular invariant and vanishes on account of theta-function identities   \cite{FarkasKra}.
 The  theta-function with standard argument represents the GSO projected contribution of the two fermions while the other theta-function is the contribution of the free boson on a circle at radius $R=\sqrt{2/3 \alpha'}$. To introduce fugacities and to exhibit the one-loop amplitude in a manifestly space-time supersymmetric form, we will exploit a generalized theta-function identity with an extra term and a non-trivial right hand side:
\begin{multline}
\theta_{3}(\nu_1,\tau) \theta_{3}( \nu_2,\frac{\tau}{3})-
\theta_{4}(\nu_1,\tau) \theta_{4}( \nu_2,\frac{\tau}{3})
-\theta_{2}(\nu_1,\tau)
\theta_{2}(\nu_2,\frac{\tau}{3}) -\theta_{1}(\nu_1,\tau) \theta_{1}(\nu_2,\frac{\tau}{3})\\ 
 = 2 \theta_{1}(\frac{1}{2} \nu_1 - \frac{3}{2} \nu_2 ,\tau) \theta_{1}(\frac{1}{2}(\nu_1+\nu_2),\frac{\tau}{3}) 
 \, .
\label{Level23Identity}
\end{multline}
It turns out that there are additional subtleties that crop up when we introduce the appropriate fugacities in the partition function. We therefore discuss this point in detail.

Our background satisfies the conditions  to have a space-time ${\cal N}=2$ superconformal symmetry \cite{Giveon:2003ku}. We thus can introduce a fugacity $\tau_{\text{s.t.}}$ for the left-moving scaling operator as well as a  fugacity $\nu$ that couples to  the boundary $R$-charge, encoded in the rotations along the space-time circle.  As before, the fugacity $\nu_1$ is identified with $\taust$, the modular parameter of the  boundary  torus. For the introduction of the fugacity $\nu$, it  is important to correctly determine the normalization of the space-time R-charge in terms of the momentum on the circle. From the analysis in  \cite{Giveon:2003ku} it transpires that at level $k=\frac23$ the space-time R-charge $Q^R_{\text{s.t.}}$ and the circle momentum are related as
\be
Q^R_{\text{s.t.}}
= 
\frac{2}{\sqrt{3}} p_L~,
\ee
where $p_L$ is the left moving momentum along the circle in the NSR formalism, in conventions where $\alpha'=2$ and the R-charge of the supercurrents is $\pm 1$. 
This implies that we must  pick the identification $\nu_2=\frac23 \nu$. 

A second simple but crucial observation is the following. There are two space-time fermionic excitations arising from the R-sector of the NSR left-movers. These two fermions pick up conformal dimensions $\pm 1/2$ as well as R-charges $\pm 1/2$. If we conserve the symmetry between negative and positive energy excitations, then the number of fermion modes is such that the spectrum inevitably breaks the charge conjugation symmetry of the space-time ${\cal N}=2$ superconformal algebra. In other words, the partition function varies under the operation $\nu \leftrightarrow -\nu$. This is seen explicitly from the fourth term on the left hand side of the theta-function identity (\ref{Level23Identity}).\footnote{There is another identity that can be obtained from this theta-function identity by charge conjugation. It gives rise to a charge conjugate choice of spectrum. An alternative to our statements is to attempt to define a theory that corresponds to the sum of charge conjugate partition functions. }
Thus, the spectrum forms a representation of   the superconformal algebra that breaks the $\mathbb{Z}_2$ outer automorphism $({\cal J},{\cal G}^{\pm}) \leftrightarrow (-{\cal J},{\cal G}^\mp)$ of the ${\cal N}=2$ superconformal algebra. When we act with this outer automorphism on our choice of partition function, we find a charge conjugate theory.

Keeping these peculiarities in mind, we next follow the well trodden path by substituting the known expression for the bosonic $AdS$ and ghost partition functions and we end up with the single string free energy in the NSNS sector of the boundary theory:
\begin{align}
f(\taust,\nu) =& \frac{1}{2 \pi} \int_0^\infty \frac{d \tau_2}{\tau_2^{\frac{3}{2}}} \int d\tau_1 \frac{e^{- \frac{2 \beta^2}{12 \pi \tau_2}}}{|\theta_1(\tau_{\text{s.t.}},\tau)|^2} 
\left|\theta_1(\frac{\taust}{2}-\nu,\tau) \theta_1\big( \frac{ \tau_{\text{s.t.}}}{2}+\frac{\nu}{3},\frac{\tau}{3}\big) \right|^2
\, . 
\end{align}
The contribution of the discrete states can  be obtained through familiar manipulations \cite{Maldacena:2000kv,Ashok:2020dnc} and the  result can be written as a sum over states:
\begin{align}
f_{\text{disc}} =& \int_{-\frac12}^{\frac12} d\tau_1\int_0^{\infty} \frac{d\tau_2}{\tau_2}\sum_{n_L, n_R f, \bar f, w, N}(-1)^{n_L+f-n_R-\bar f} \int_{(\frac12, \frac{5}{6}]}  \frac{dj}{\pi}
q^{L_0 -\frac12} \bar q^{\bar L_0 -\frac12}\cr
&\hspace{0.1cm}
d(j,r,f,n_{L,R},q) \, 
\qst^{\frac{w}{3} + j+r+\frac{n_L+f-1}{2}} \qbarst^{\frac{w}{3}  + j + \bar r+\frac{n_L+\bar f-1}{2}}~
~z^{-\frac{2w}{3}-f+\frac{n_L}{3}+\frac13}~\bar z^{-\frac{2w}{3}-\bar f+\frac{n_R}{3}+\frac13}~.
\label{fdisc23}
\end{align}
The left-moving worldsheet Virasoro generator is given by
\begin{align}
L_0&=  -\frac{3}{2}j(j-1)-w\big(j+r+\frac23(n_L-\frac12)\big) + \frac16(n_L-\frac12)^2 + \frac12(f-\frac12)^2+N~,
\end{align}
and similarly for the right-movers.

\subsection{States of Minimal Dimension}
We first study the spectrum of single string excitations in the NSNS sector since it will already reveal an interesting feature of the theory. Subsequently we  will study the RR sector.

\subsubsection{Low Lying States in the NSNS sector}
The left-moving conformal dimension and R-charge in space-time in terms of the world sheet quantum numbers of the string are, according to equation (\ref{fdisc23}):
\be
H = \frac{w}{3}+j+r+\frac{1}{2}(n_L+f-1)~,\qquad Q^R_{\text{s.t.}} = -\frac{2w}{3}-f+\frac{n_L}{3}+\frac13~.
\ee
Our first goal is to find the state with the lowest conformal dimension $H$. For this purpose it is optimal to set $r=0$ and to have no oscillators excited. In the $w=0$ sector,  the lowest conformal dimension is obtained for the choices $n_L=f=0$, which corresponds to spin $j=\frac23$ and dimension $H=\frac16$.  This state has $Q^R_{\text{s.t.}} = \frac{1}{3}$ and it is therefore a chiral primary state. It was described in \cite{Giveon:1999zm}. Using our results, one can prove that this is  the state with the minimum space-time conformal dimension among all discrete states, including those with $w \ge 1$. 

We turn to  the states from the continuous part of the world sheet spectrum and write down the conformal dimension and worldsheet generators for a state with Casimir labelled by  $j=\frac12+ \ii s$ and whose left-moving $\mathfrak{u}(1) \subset \mathfrak{sl}(2)$ eigenvalue is given by the real number $m$ \cite{Maldacena:2000hw}. For these excitations, we have the space-time and world sheet conformal dimensions:
\begin{align}
H & = \frac{w}{3} + m+\frac{1}{2}(n_L+f-1)\\
L_0&= -\frac32j(j-1)- w(m+\frac23(n_L-\frac12)) + \frac16(n_L-\frac12)^2+\frac12(f-\frac12)^2~.
\end{align}
We first solve for the left-moving $\mathfrak{u}(1)$ time-like momentum $m$ using the on-shell condition $L_0=\frac12$ and substitute it into the expression for $H$:
\be
H= \frac{1}{6w}\left(n_L-\frac{w+1}{2} \right)^2 + \frac{1}{2w}\left(f-\frac{(1-w)}{2}\right)^2+ \frac{w}{6} +\frac{1}{w}\left(\frac{3s^2}{2} - \frac{1}{8}\right)~.
\ee
One can choose the optimal values $n_L = \left\lfloor {\frac{w+1}{2}} \right\rfloor $, $f =\left \lfloor {\frac{1-w}{2}}\right\rfloor$ and $s=0$, to find that 
\be
H_{\text{min}}(w) =
\begin{cases}
 \frac{w}{6} - \frac{1}{8w}  = \frac{1}{6}(w-\frac{1}{w}) + \frac{1}{24w}& \text{ for odd $w$}\\
 \frac{w}{6} + \frac{1}{24w} = \frac{1}{6}(w-\frac{1}{w}) + \frac{5}{24w}& \text{ for even $w$} \, .
\end{cases}
\ee
This form of the conformal dimension agrees  for odd $w$ with the results obtained  in \cite{Giveon:2005mi}. The alternate form for even $w$ is due to the GSO projection.

We come to a first important conclusion.
The global minimal conformal dimension over all winding sectors in the space-time NSNS sector equals $H_\text{min}=1/24$ and the minimum  is reached at the bottom of the continuous representations for a singly wound fundamental string. The R-charge of the state is $Q^R_{\text{s.t.}}=0$.
The minimal conformal dimension is strictly positive. 
Therefore there is no normalizable $SL(2,\mathbb{C})$ invariant ground state \cite{Giveon:2005mi}.\footnote{The spontaneous breaking of conformal invariance is  a property
of Liouville theory, and therefore also of its speculative three-dimensional gravitational dual \cite{Krasnov:2000zq,Li:2019mwb}. } The lowest lying chiral primary state at $H=1/6$  partially breaks supersymmetry. The ground state at zero charge breaks supersymmetry entirely.

\subsubsection{The Lowest Lying Ramond-Ramond States}
Recall that our partition function breaks charge conjugation symmetry.
There are therefore two inequivalent directions in which we can flow from the single string NSNS sector to Ramond-Ramond sector states.
To find RR sector space-time conformal dimensions, we either add or subtract the R-charge. 
We first discuss what happens when we flow such that the Ramond-Ramond left-moving conformal dimension gap above the ground states equals
\begin{equation}
H_{RR} = \frac{2w}{3}+j+r+f+\frac{n_L}{3}-\frac{2}{3} \, .
\end{equation} 
The only RR ground states with $H_{RR}=0$ are found at the quantum numbers $(j,n_L,f,w)=(2/3,w,-w,w)$ with R-charges $Q= (2w+1)/3$ and $w \ge 0$.\footnote{We note that the R-charge formula for the Ramond-Ramond ground states is of the form $Q=2j-1+kw$ as in other instances.} The spectrum of R-charges is infinite and fractional.
In the continuous sector, the gap to the Ramond-Ramond ground states is $1/(24 w)$. It is proportional to one over the winding and closes at large winding.
The second way of spectrally flowing the single string states to the  Ramond-Ramond sector gives rise to no extra Ramond-Ramond ground states.
The two lowest lying discrete states that one obtains in this manner have zero winding and  dimension  $H_{RR}=1/3$.
The lowest lying states in the continuous sector again have a gap of $1/(24w)$ with four-fold degeneracy.

\section{The Thermodynamics Beyond the String Scale}
\label{BeyondStringScale}
For our discussion of the finite temperature thermodynamics of the non-critical $AdS_3$ string theories beyond the string scale, it will be useful to have models at hand that have properties that interpolate between those of our two main examples at levels $k=1$ and $k=2/3$. 

\subsection{An Interpolation}
To that end, we study the models of the form \cite{Giveon:1999zm}:
\begin{equation}
AdS_3 \times S^1 \times \text{Minimal Model} \, 
\end{equation}
where the $N=2$ superconformal minimal model is of central charge $c_{\text{MM}} = 3-6/n$. Therefore, the supersymmetric $AdS_3$ model is at level $k=n/(n+1)$. The model still falls in the class of backgrounds which permit a ${\cal N}=2$ superconformal algebra in space-time with a $U(1)_R$ symmetry associated to the circle direction.  We revisit two properties of the model: the first is the set of chiral primaries in space-time and the second is the minimal conformal dimension in the continuum in the space-time NSNS sector. See also \cite{Giveon:1999zm,Argurio:2000tb,Giveon:2005mi}. 

We work in the NSR formalism\footnote{It should be possible to formulate this model in the Green-Schwarz formalism as well. This requires deriving suitable vanishing identities involving theta functions and characters of minimal models.} in which the world sheet conformal dimension $L_0$, space-time conformal dimension $H$ and space-time R-charge $Q^R_{\text{s.t.}} $ read:
\begin{align}
L_0 &= L_0^w + L_0^{S^1} + L_0^{{\text{Min Mod}}} \cr
&= - \frac{j(j-1)}{k}-w (j+r)- \frac{k+2}{4} w^2 + \frac{(f+a)^2}{2} + + L_0^{S^1} + L_0^{{\text{Min Mod}}} \, ,
\\
H &= j+r+ f +a + \frac{k+2}{2} w \, ,\\
Q^R_{\text{s.t.}} &= \sqrt{2k} p_L 
%= \sqrt{2n/(n+2)} (n/\sqrt{2k}+w \sqrt{2k}/2) 
=  m' + \frac{n}{n+1} w'  \, .
\end{align}
Our conventions are such that $a =0$ for the worldsheet NS sector and $a=-1/2$ for the  R sector. The worldsheet fermion number in the supersymmetric $AdS_3$ sector is denoted $f$ and the quantum numbers $(m', w')$ correspond to the momentum and winding along the $S^1$ direction. 

\subsubsection{Space-time Chiral Primaries}

We look for chiral primaries in the space-time NSNS sector that satisfy the lower bound $H=Q^R_{\text{s.t.}}/2$. In principle these can arise from either the NS or R worldsheet sectors. In this particular example, it turns out that all boundary chiral primaries arise from  the worldsheet RR sector only. 
Inspired by the solutions found for the $n=2$ case,  we look for solutions of the form  
\be
H=j-\frac12+\frac{kw}{2}~,
\ee
by setting $f=-w$ and $a=-1/2$ which implies that we are in the worldsheet R sector. We also set $r=0$. It is key to find the remaining quantum numbers such that we satisfy the on-shell condition $L_0=\frac12$. We recognize that the contribution to the world sheet conformal dimension $L_0$ from the circle sector is given directly in terms of the space-time R-charge:
\be
L_0^{S^1} = \frac{p_L^2}{2} = \frac{\big(Q^R_{\text{s.t.}}\big)^2}{4k} = \frac{1}{k}\big(j-\frac12+\frac{kw}{2}\big)^2~.
\ee
Rearranging the terms in the worldsheet conformal dimension we find that it can be written in the  suggestive form:
\begin{align}
L_0 
&=\frac{1}{4k}+\frac18 +L_0^{{\text{Min Mod}}}~.
\end{align}
The minimal model contribution  from the world sheet Ramond sector is at least
\be
L_0^{{\text{Min Mod}}} = \frac{c_{\text{MM}}}{24}=\frac{1}{8} - \frac{1}{4n}~.
\ee
Substituting this into the expression for the worldsheet conformal dimension and using the fact that $\frac1k = 1+\frac1n$, we find that $L_0=\frac12$
precisely for the minimal model ground states. It  remains to plug these quantum numbers into the expression for the space-time R-charge and find the $S^1$ quantum numbers $(m',w')$ as well as the allowed values of the spin $j$. From the saturation of the space-time chiral primary bound, we have to look for integers $(m', n')$ that correspond to momentum and winding quantum numbers on the $S^1$, such that
\be
m' + \frac{n}{n+1} w' = 2j-1+ \frac{n}{n+1}w~.
\ee
There is a unique solution (up to equivalences) for every value of the spectral flow number $w$. Before we present our result, it is important to recall that the $AdS_3$ spin $j$ takes values in the half open interval $ j \in \big(\frac12, \frac{2n+1}{2(n+1)}\big]$ in the discrete sector. 
Then the general solution for $(j, m', n')$ can be written in terms of a single integer $p$: 
\begin{align}
\label{allowedj}
j &= \frac12+\frac{p}{2(n+1)}~,
\qquad m' = p~, \qquad w' = w -p~\quad \text{for} \quad p = 1,2, \ldots n~.
\end{align}
In the NSR formalism, we must GSO project which  amounts to imposing that the world sheet R-charges be odd in both the left and right moving sectors. 
The world sheet R-charge $Q_{ws}$ and the GSO constraint   are:
\be
Q_{ws} = \frac{s_0+s_1}{2}-\frac{m}{n}+ \frac{n+1}{n} m' + w' \in \text{odd integers} \, ,
\ee
where $s_0$ and $s_1$ are fermionic quantum numbers corresponding respectively to the pair of $AdS_3$ fermions and the minimal model, and $m$ labels the minimal model spin component. The minimal model Ramond sector ground states are labelled by spins $\ell= 0, \frac 12, \ldots \frac{n}{2}-1$ and have spin components $m=2\ell+1$ as well as $s_1=1$. 
From the allowed values of $(j,m',n')$ in equation \eqref{allowedj}, it follows that the constraint on the worldsheet R-charge is given by
\be
Q_{ws} = \frac{s_0+1}{2}+\frac{p-(2\ell+1)}{n}+ w \in \text{odd integers}.
\ee
We note that $2\ell+1$ runs over integers in the range $[1, n-1]$. Accordingly this leads to an omission of the $p=n$ state from the allowed set of states as for this state, the fractional contribution to the worldsheet R-charge proportional to $1/n$ cannot be cancelled by any choice of the spin $\ell$. One can then choose the sign of the fermion number $s_0$  such that for any winding $w$, one has one state that survives the GSO projection. We have thus shown that for a given winding $w$ there are $n-1$ chiral primary states in the boundary theory with conformal dimension equal to 
\be
H = \frac{p + n w}{2(n+1)}~, \quad \text{for}\quad p = 1, 2, \ldots n-1~.
\ee
This is in accord with results that can be extracted from   \cite{Giveon:1999zm,Giveon:2005mi, Argurio:2000tb}.

\subsubsection{Continuum modes}

Finally, we review the minimal conformal dimension in the continuum in the NSNS sector for each value of the winding $w \ge 1$ \cite{Seiberg:1999xz,Giveon:2005mi}. The world sheet constraint equation and the space-time energy in the world sheet NSNS sector (on which we concentrate) are of the form:
\begin{align}
\label{L0forcontk}
L_0 &= L_0^w + L_0^{S^1} + L_0^{MM} \nonumber \\
&= - \frac{j(j-1)}{k}-w m- \frac{k+2}{4} w^2 + \frac{f^2}{2} + + L_0^{S^1} + L_0^{{\text{Min Mod}}} \, ,\\
H &= m+ f  + \frac{k+2}{2} w \, .
\end{align}
We use the on-shell condition $L_0=\frac12$ to solve for $m$ and substitute into the expression for the conformal dimension $H$. Extremizing with respect to $(f, p_L, L_0^{{\text{Min Mod}}})$ we find that the optimal configuration to minimize the space-time conformal dimension $H$ is to set $f=-w$ and for the $S^1$ and minimal model to be in their ground state.  The minimal conformal dimension as a function of the winding $w$ is
\begin{equation}
H_{NSNS}^{\text{min}} (w) = \frac{H_{NSNS}^{\text{min}}}{w} + \frac{k}{4} (w-\frac{1}{w}) \, ,
\end{equation}
where the minimal  conformal dimension for the singly wound long string is:
\begin{equation}
H_{NSNS}^{\text{min}} = \frac{(k-1)^2}{4k} \, . \label{MinimalConformalDimension}
\end{equation}
Indeed, the state with $w=1$ satisfies the GSO projection. 
This result agrees with \cite{Seiberg:1999xz,Giveon:2005mi}.
This provides the global minimum of the NSNS sector conformal dimension for the $k<1$ string theories. In the previous section, we have proven this more rigorously for the minimal $k=2/3$ theory. 
Finally, we note that this class of theories also breaks charge conjugation symmetry spontaneously, by the same mechanism described in detail for the theory at level $k=2/3$. 

It is useful to have the models in this section in mind as examples for statements in the next section.

\subsection{The Thermodynamics}
In this section, we recall an analysis of the entropy, free energy and critical temperatures of thermal $AdS_3$ super string theory and describe how the particular backgrounds that we studied fit into the big picture. We add a few remarks on the finite temperature thermodynamics beyond the string scale.

\subsubsection{The Entropy and The Correspondence Point}
For curvature radii below the string scale, $k<1$, neither the $AdS_3$ ground state nor the S-dual BTZ black hole states are normalizable \cite{Giveon:2005mi}. At these extreme values of the curvature, the minimal conformal dimension for operators in the (space-time) NSNS sector is positive. Therefore, the effective central charge which counts the number of degrees of freedom deviates from its bare value, and we have a qualitative change in the behaviour of the entropy at the correspondence point $k=1$
\cite{Horowitz:1996nw,Giveon:2005mi}.  It is interesting to study how the transition to the case $k \ge 1$ behaves.
We recall that the  background bare central charge $c^{\text{s.t.}}$ in space-time is $c^{\text{s.t.}}= 6 k Q_1$ where $Q_1$ is the background number of fundamental strings. The background number of fundamental strings is proportional to $1/g_s^2$ and is therefore infinite in perturbation theory. It can be viewed as setting a large cut-off on the number of winding strings $w$ that can be produced. 
The effective central charge when the minimal conformal dimension is positive is $c_{\text{eff}}^{\text{s.t.}}=c^{\text{s.t.}}-24 H_{NSNS}^{\text{min}}$ \cite{Giveon:2005mi}. The effective central charge captures the entropy of the modes of high conformal dimension. The central charge has a discontinuous second derivative at the correspondence point $k=1$ -- see formula (\ref{MinimalConformalDimension}). We note that the curvature radius $k$ is fixed by the asymptotic boundary conditions on our quantum theory of gravity and that therefore we described the qualitatively different behaviour of distinct theories.

\subsubsection{The Free Energy and Critical Temperatures}
The correspondence point is associated to an analysis of the entropy of the boundary conformal field theory \cite{Giveon:2005mi}. We can draw further conclusions on the phases of $AdS_3$ string theory by studying the free energy of the  background at finite temperature. The first observation is that at finite temperature, there is a Hawking-Page transition between the thermal $AdS_3$ background and the dual BTZ black hole. The
Hawking-Page temperature equals $T_{HP}^{(2)}= (2 \pi)^{-1}$ in string units \cite{Hawking:1982dh,Witten:1998zw,Maldacena:1998bw}. Above this temperature, the BTZ black hole is thermodynamically favoured over thermal $AdS_3$, and the gas of gravitons collapses into a black hole. The calculation of the phase transition has  been done in terms of the metric -- the thermodynamics of quantum gravity is captured by general relativity to leading order. In Appendix \ref{HPWS}, we provide a derivation of the Hawking-Page transition that directly uses fundamental string perturbation theory and a non-zero $AdS_3$ one-point function. 

Further interesting phase transitions are coded in the spectrum of fundamental string fluctuations around a given classical background. The inverse Hagedorn temperature $\beta_H$ for fundamental super strings in the $AdS_3$ background  can be derived by applying the reasoning of \cite{Berkooz:2007fe} to the super string amplitude. We concentrate on the world sheet NSNS sector and recall the unfolded one-loop amplitude:
\be
\begin{aligned}
Z_{1-loop}(\beta, \mu) &= \frac{\beta\sqrt{k}}{8\pi} \int_0^\infty \frac{d \tau_2}{\tau_2^{3/2}} \int_{-\frac{1}{2}}^{\frac{1}{2}} d \tau_1 
\sum_{m=1}^{\infty} \frac{ e^{-\frac{k m^2\beta^2} { 4 \pi \tau_2}} }{|\theta_1(  m\tau_{\text{s.t.}},\tau )|^2} ~ {Z}_{int} (q) .
\end{aligned}
\ee
The world sheet partition function $Z_{\text{int}}$ accounts for all degrees of freedom except the $AdS_3$ bosons. In the superstring setting, the  central charge of the internal partition function is $c_{int}=12-(3+6/k)$. Modular invariance determines the  small $\tau_2$ behaviour of the integrand:
\begin{equation}
e^{- \frac{k \beta^2}{4 \pi \tau_2}}~e^{ \frac{4 \pi (12- \frac{6}{k})}{24 \tau_2}}~.
\end{equation}
That leads to the critical Hagedorn temperature at which the exponential degeneracy of states swamps the  thermal suppression factor:
\begin{equation}
\beta_H = \frac{2 \pi}{\sqrt{k}} \big(2 - \frac{1}{k}\big)^{\frac{1}{2}} \, .
\end{equation}
This corresponds to the temperature:
\begin{equation}
T_H^{(3)} = \frac{\sqrt{k}}{2 \pi}\big (2 - \frac{1}{k}\big)^{-\frac{1}{2}}
\end{equation}
We also have a similar Hagedorn transition in the dual BTZ background -- the path integrals are identical up to the boundary modular parameter --:
\begin{equation}
T_H^{(1)} = \frac{1}{2 \pi \sqrt{k}} (2 - \frac{1}{k})^{\frac{1}{2}} \, .
\end{equation}
When the level $k>1$, the critical temperatures arrange themselves in the order $T_{H}^{(1)}<T_{HP}^{(2)}<T_{H}^{(3)}$. 
The resulting thermodynamic phases were discussed  in
\cite{Berkooz:2007fe}. At the correspondence point $k=1$, all the critical  temperatures are equal. The Hagedorn and the Hawking-Page transition become indistinguishable. %The overlap of the validity of the BTZ and the AdS background narrows to a point.

Beyond the correspondence curvature, the naive hierarchy of temperatures is re-established in the original order. Indeed, {\em if} we put trust in our critical temperature formulas for $k<1$, we again have the order $T_{H}^{(1)}<T_{HP}^{(2)}<T_{H}^{(3)}$. Thus, the description of the phases would be as in the $k>1$ regime. This is explained by the amusing duality\footnote{In the bosonic string theory, a similar duality holds.}
\begin{equation}
k_{\text{dual}} = \frac{k}{2k-1} \, ,
\end{equation}
that leaves the critical temperatures invariant.
The critical, self-dual point is $k_{sd
}=1$. Thus, the correspondence point is also special from the perspective of the Hagedorn transition.  We should stress that these are speculations because they are based on formulas that are computed by expanding around a non-normalizable state.
It may  be that it is ill advised to do so
\cite{Berkooz:2007fe}. Still, the one-loop amplitude  only exhibits the same volume divergences below string radius as it does above.

\section{Conclusions}
\label{Conclusions}
We studied thermal three-dimensional anti-de Sitter backgrounds that arise in non-critical string theories near a density of fundamental strings. These $AdS_3$ backgrounds provide  examples of holography that can be just above, at, or even below the string scale. They are ideal to uncover strong curvature effects in string theory holography. The one loop vacuum amplitude or free energy of the superstring theories has an integrand that (almost) factorizes into partition functions of the world sheet conformal field theories. One factor  is the partition function of the non-compact interacting conformal field theory on $AdS_3$ with NSNS flux, while the remaining partition functions take a form perfectly suited to the application of generalized Jacobi identities. These identities allow us to move between a NSR description of the background and a manifestly supersymmetric description,  including fugacities that keep track of the space-time conformal dimensions and R-charges of all single string states. The path integral approach thus provides  a complete and efficient description of the  spectrum that allows for classification results.

We analyzed a few properties of these string scale  holographic theories. We determined the single string free energy and classified the supersymmetric Ramond-Ramond ground states. In particular, we studied a ${\cal N}=2$ superconformal theory precisely at the string scale. The second quantized Ramond-Ramond ground state spectrum matches that of a symmetric orbifold of a two-torus. Notably, our theory at the correspondence point has  a continuous part to its spectrum.
We also studied a background which is naturally at a scale below the string scale. In the NSNS sector, we showed that there is a spontaneous breaking of both the conformal symmetry, supersymmetry as well as a charge conjugation symmetry. In the Ramond-Ramond sector we determined an infinite set of ground states with fractional R-charges. We made speculative remarks on the thermodynamics of these string backgrounds at very strong curvature. 

These string scale non-critical $AdS_3$ backgrounds, their topological twists and their holographic duals deserve further study. For instance, there are many techniques to solve for correlation functions that hinge on having the critical level $k=1$, and that are bound to be applicable to all theories at the correspondence point. It will be interesting to understand the effect  of the presence of the continuous part of the spectrum on the three-point correlation functions. 

In an appendix, we showed how to derive the Hawking-Page transition from the point of view  of the theory on the perturbative fundamental string. The zero-point function is determined by integrating a non-zero one-point function for the central charge operator. The resulting space-time effective action  correctly captures the tree level $AdS_3$ background thermodynamics. 

There is a zoo of $AdS_3$ string theories both above, at, and below the string scale. We scrutinized a few quintessential examples from the perspective of their one loop free energy and laid bare interesting properties of their spectra. In this large collection of theories, a number of crucial observables remains to be computed and each may shed more light on the original properties of their holographic duals, which remain odd ducks in the family of two-dimensional conformal field theories.

\appendix

\section{The Hawking-Page Transition from the World Sheet }
\label{HPWS}
In this appendix, we derive the Hawking-Page transition from perturbative world sheet string theory. 
\subsection{The Gravitational Analysis}
We recall that the Hawking-Page transition manifests itself in the free energy of the thermal $AdS_3$ and euclidean BTZ background as calculated in euclidean gravity \cite{Hawking:1982dh,Witten:1998zw,Maldacena:1998bw}. The on-shell classical action of general relativity $I$ is given by:
\begin{equation}
I = - \frac{i \pi c}{12} (\tau - \bar\tau) \, , \label{InstantonAction}
\end{equation}
where $\tau$ is the modular parameter associated to the background. If for euclidean $AdS_3$ it is $\tau=\taust$, defined in \eqref{boundarytau} then for its S-dual BTZ background it is $\tilde{\tau}=-1/\tau$.
When we consider zero angular chemical potential and concentrate on $\tau=\ii \tau_2=\ii \frac{\beta}{2 \pi}$, the minimal free energy $F=I/\beta$ is attained by either thermal $AdS_3$ or the black hole, with a   Hawking-Page transition point at $\tau_2=1$ or $\beta=\beta_{HP}= 2\pi$ \cite{Hawking:1982dh,Witten:1998zw,Maldacena:1998bw}.

\subsection{The Fundamental String's Point of View}
The Hawking-Page transition is uncovered from the classical on-shell gravitational action. It is therefore a string theory tree level effect, and will  be associated to a spherical world sheet $\Sigma=S^2$. 
To determine the space-time action from string theory, we use the idea that we can identify the tree level closed string space-time effective action with the world sheet spherical partition function. See \cite{Tseytlin:1988rr} for a review and references.
 Moreover, rather than derive the action directly, we will derive its dependence on the space-time boundary modular parameter $\tau$ through variation, and then integrate the result up.
Indeed, in gravity as well, we can derive the dependence of the euclidean instanton action $I$ on the boundary modular parameter $\tau$ via the dependence of the gravity action on the boundary metric, captured by the energy-momentum tensor.

 Thus, inspired by \cite{Tseytlin:1988rr},
we start out as in \cite{Giveon:1998ns,Kutasov:1999xu,Li:2020nei}. We have the sigma-model partition function $Z$ which is a function of the boundary metric $g^{(0)}$ on the asymptotically anti-de Sitter space:
\begin{equation}
Z[g^{(0)}] = \frac{1}{g_s^2} Z_{\text{sphere}}[g^{(0)}] + \dots
\end{equation}
where we concentrate on the tree level term that arises from the partition function for a spherical string world sheet $\Sigma=S^2$.
We work with a euclidean anti-de sitter space-time with torus boundary of modular parameter $\tau$. We have the asymptotic metric:
\begin{equation}
d s^2 = k \alpha' (d \phi^2 + e^{2 \phi} d x d \bar{x}) \, 
\end{equation}
with the toroidal identification $x \equiv x + 2 \pi \equiv x + 2 \pi \tau$.  
We  compute the variation of the space-time action as a function of the boundary complex structure, at fixed volume of the two-torus. 
The boundary metric then undergoes the variations:
\begin{equation}
h_{\bar x \bar{x}} = \delta g_{\bar{x} \bar{x}}^{(0)}= -\frac{\delta \tau}{\tau-\bar{\tau}}
\qquad
h_{x {x}} =\delta g_{x x}^{(0)} = \frac{\delta \bar{\tau}}{\tau - \bar{\tau}}
\qquad
    h_{x \bar{x}} =
\delta g_{x \bar{x}}^{(0)} =  0
\, .
\end{equation}
We define the space-time boundary energy momentum tensor ${\cal T}$ as the derivative of the functional integral $Z = \int dX e^{-S}$ with respect to the boundary metric:
\begin{equation}
{\cal T}^{\alpha \beta} =  \frac{ 4 \pi \delta S}{\sqrt{g^{(0)}} \delta h_{\alpha \beta}} \, .
\end{equation}
The variation of the world sheet  partition function is:
\begin{equation}
\delta Z = - \langle \delta S \rangle
=  -i \int_{T^2} \frac{\sqrt{g^{(0)}}}{4 \pi \tau_2} ( \delta \tau
\langle {\cal T}^{\bar{x} \bar{x}} \rangle -\delta \bar{\tau} \langle {\cal T}^{{x} {x}} \rangle) \, .  \label{VariationAction}
\end{equation}
The boundary metric determinant as well as the variations of the complex structure are independent of the world sheet fields. We have introduced the world sheet energy-momentum component vertex operators, which can be found explicitly in \cite{Giveon:1998ns,Kutasov:1999xu,Li:2020nei}.

We need to evaluate the one-point function $\langle {\cal T}_{xx} \rangle$ when the boundary manifold is a torus.  For that boundary topology,  the one-point function of the energy-momentum tensor component is proportional to the one-point function of the central charge operator $C$ that is part of the energy-momentum tensor operator product algebra \cite{Giveon:1998ns,Kutasov:1999xu}.
We  have:
\begin{equation}
\langle {\cal T}_{xx} \rangle = \frac{\langle C \rangle}{24}  \, . \label{EnergyMomentum}
\end{equation}
A crucial step is the world sheet evaluation of the 
central charge one-point function $\langle C \rangle$ which equals \cite{Troost:2011ud}:
\begin{equation}
\langle C \rangle 
=c % \, \, \frac{\log e^{\phi_{IR}}}{\log \epsilon^{-1}}
\label{CentralCharge}
\end{equation}
where $c= \frac{3}{2} \frac{\sqrt{k \alpha'}}{G_N^{(3)}}$ is the $\alpha'$ exact value of the Brown-Henneaux central charge \cite{Brown:1986nw} in the supersymmetric theory.
We combine the results (\ref{VariationAction}), (\ref{EnergyMomentum}) and (\ref{CentralCharge}) and integrate the world sheet partition function $Z$ which we identify with the space-time instanton action $I$:
\begin{equation}
I=Z = - i \pi \frac{c}{12} (\tau - \bar{\tau}) 
%\frac{\log e^{\phi_{IR}}}{\log \epsilon^{-1}} 
\, .
\label{Identification}
\end{equation}
As reviewed above, this result and its  $SL(2,\mathbb{Z})$ duals (which are derived identically) code the Hawking-Page transition. 

In flat space backgrounds, there is ground laying literature  identifying the space-time closed string effective action with the logarithmic derivative of the world sheet partition function with respect to the world sheet ultraviolet cut-off $\epsilon$ \cite{Tseytlin:1988rr}. It is worth remarking that in our calculation above, a similar feature is implicit. In  three-dimensional anti-de Sitter space the space-time has a volume proportional to the volume of the conformal group and we  may have a cancellation between space-time and world sheet divergences. The divergences 
in the central charge one-point function are milder, but indeed, in the calculation of the one-point function, a space-time logarithmic infrared divergence is cancelled against a world sheet logarithmic ultraviolet divergence  \cite{Troost:2011ud}. See \cite{Maldacena:2001km} as well as \cite{Buchbinder:2011jr} for discussions of the generic mechanism. This allows for the seemingly naive identification of the world sheet partition function and the space-time action (\ref{Identification}), reminiscent of a similar identification in open string theory in flat space \cite{Tseytlin:1988rr}.

\bibliographystyle{JHEP}

\begin{thebibliography}{99}

%\cite{tHooft:1993dmi}\cite{Susskind:1994vu}
\bibitem{tHooft:1993dmi}
G.~'t Hooft,
``Dimensional reduction in quantum gravity,''
Conf. Proc. C \textbf{930308} (1993), 284-296
[arXiv:gr-qc/9310026 [gr-qc]].
%2468 citations counted in INSPIRE as of 27 Jan 2021

%\cite{Susskind:1994vu}
\bibitem{Susskind:1994vu}
L.~Susskind,
``The World as a hologram,''
J. Math. Phys. \textbf{36} (1995), 6377-6396
doi:10.1063/1.531249
[arXiv:hep-th/9409089 [hep-th]].
%3030 citations counted in INSPIRE as of 27 Jan 2021

%\cite{Maldacena:1997re}
\bibitem{Maldacena:1997re}
J.~M.~Maldacena,
``The Large N limit of superconformal field theories and supergravity,''
Int. J. Theor. Phys. \textbf{38} (1999), 1113-1133
doi:10.1023/A:1026654312961
[arXiv:hep-th/9711200 [hep-th]].
%16386 citations counted in INSPIRE as of 27 Jan 2021

%\cite{Eberhardt:2019ywk}\cite{Giribet:2018ada}
\bibitem{Eberhardt:2019ywk}
L.~Eberhardt, M.~R.~Gaberdiel and R.~Gopakumar,
``Deriving the AdS$_{3}$/CFT$_{2}$ correspondence,''
JHEP \textbf{02} (2020), 136
doi:10.1007/JHEP02(2020)136
[arXiv:1911.00378 [hep-th]].
%32 citations counted in INSPIRE as of 27 Jan 2021


%\cite{Giribet:2018ada}
\bibitem{Giribet:2018ada}
G.~Giribet, C.~Hull, M.~Kleban, M.~Porrati and E.~Rabinovici,
``Superstrings on AdS$_{3}$ at ${k} =$ 1,''
JHEP \textbf{08} (2018), 204
doi:10.1007/JHEP08(2018)204
[arXiv:1803.04420 [hep-th]].
%49 citations counted in INSPIRE as of 27 Jan 2021


%\cite{Giveon:2005mi}
\bibitem{Giveon:2005mi}
A.~Giveon, D.~Kutasov, E.~Rabinovici and A.~Sever,
``Phases of quantum gravity in AdS(3) and linear dilaton backgrounds,''
Nucl. Phys. B \textbf{719} (2005), 3-34
doi:10.1016/j.nuclphysb.2005.04.015
[arXiv:hep-th/0503121 [hep-th]].
%79 citations counted in INSPIRE as of 27 Jan 2021

%\cite{Hawking:1982dh}
\bibitem{Hawking:1982dh}
S.~W.~Hawking and D.~N.~Page,
``Thermodynamics of Black Holes in anti-De Sitter Space,''
Commun. Math. Phys. \textbf{87} (1983), 577
doi:10.1007/BF01208266.
%1875 citations counted in INSPIRE as of 29 Jan 2021

%\cite{Hawking:1982dh,Witten:1998zw,Maldacena:1998bw}
\bibitem{Witten:1998zw}
E.~Witten,
``Anti-de Sitter space, thermal phase transition, and confinement in gauge theories,''
Adv. Theor. Math. Phys. \textbf{2} (1998), 505-532
doi:10.4310/ATMP.1998.v2.n3.a3
[arXiv:hep-th/9803131 [hep-th]].
%3185 citations counted in INSPIRE as of 27 Jan 2021


%\cite{Maldacena:1998bw}
\bibitem{Maldacena:1998bw}
J.~M.~Maldacena and A.~Strominger,
``AdS(3) black holes and a stringy exclusion principle,''
JHEP \textbf{12} (1998), 005
doi:10.1088/1126-6708/1998/12/005
[arXiv:hep-th/9804085 [hep-th]].
% %706 citations counted in INSPIRE as of 22 Oct 2020


%\cite{Berkooz:2007fe}
\bibitem{Berkooz:2007fe}
M.~Berkooz, Z.~Komargodski and D.~Reichmann,
``Thermal AdS(3), BTZ and competing winding modes condensation,''
JHEP \textbf{12} (2007), 020
doi:10.1088/1126-6708/2007/12/020
[arXiv:0706.0610 [hep-th]].
%15 citations counted in INSPIRE as of 27 Jan 2021


%\cite{Horowitz:1996nw}
\bibitem{Horowitz:1996nw}
G.~T.~Horowitz and J.~Polchinski,
``A Correspondence principle for black holes and strings,''
Phys. Rev. D \textbf{55} (1997), 6189-6197
doi:10.1103/PhysRevD.55.6189
[arXiv:hep-th/9612146 [hep-th]].
%551 citations counted in INSPIRE as of 27 Jan 2021


%\cite{Bilal:1986ia,Bilal:1986uh,Mizoguchi:2000kk,Murthy:2003es}
\bibitem{Bilal:1986ia}
A.~Bilal and J.~L.~Gervais,
``Modular Invariance for Closed Strings at the New Critical Dimensions,''
Phys. Lett. B \textbf{187} (1987), 39-44
doi:10.1016/0370-2693(87)90068-2.
%43 citations counted in INSPIRE as of 12 Jan 2021
%\cite{Bilal:1986uh}\cite{Mizoguchi:2000kk,Murthy:2003es}
\bibitem{Bilal:1986uh}
A.~Bilal and J.~L.~Gervais,
``New Critical Dimensions for String Theories,''
Nucl. Phys. B \textbf{284} (1987), 397-422
doi:10.1016/0550-3213(87)90042-3.
%63 citations counted in INSPIRE as of 12 Jan 2021

%\cite{Mizoguchi:2000kk,Murthy:2003es}

\bibitem{Mizoguchi:2000kk}
S.~Mizoguchi,
``Modular invariant critical superstrings on four-dimensional Minkowski space times two-dimensional black hole,''
JHEP \textbf{04} (2000), 014
doi:10.1088/1126-6708/2000/04/014
[arXiv:hep-th/0003053 [hep-th]].
%36 citations counted in INSPIRE as of 18 Dec 2020

%\cite{Murthy:2003es}
\bibitem{Murthy:2003es}
S.~Murthy,
``Notes on noncritical superstrings in various dimensions,''
JHEP \textbf{11} (2003), 056
doi:10.1088/1126-6708/2003/11/056
[arXiv:hep-th/0305197 [hep-th]].
%60 citations counted in INSPIRE as of 18 Dec 2020

%\cite{Giveon:1999zm}\cite{Argurio:2000tb}
\bibitem{Giveon:1999zm}
A.~Giveon, D.~Kutasov and O.~Pelc,
``Holography for noncritical superstrings,''
JHEP \textbf{10} (1999), 035
doi:10.1088/1126-6708/1999/10/035
[arXiv:hep-th/9907178 [hep-th]].
%182 citations counted in INSPIRE as of 09 Feb 2021



%\cite{Maldacena:2000kv}
\bibitem{Maldacena:2000kv}
J.~M.~Maldacena, H.~Ooguri and J.~Son,
``Strings in AdS(3) and the SL(2,R) WZW model. Part 2. Euclidean black hole,''
J. Math. Phys. \textbf{42} (2001), 2961-2977
doi:10.1063/1.1377039
[arXiv:hep-th/0005183 [hep-th]].
%240 citations counted in INSPIRE as of 27 Jan 2021


%\cite{Polchinski:1998rq}
\bibitem{Polchinski:1998rq}
J.~Polchinski,
``String theory. Vol. 1: An introduction to the bosonic string,'' Cambridge University Press, 1998 
doi:10.1017/CBO9780511816079.
%469 citations counted in INSPIRE as of 02 Nov 2020

%\cite{Ashok:2020dnc}
\bibitem{Ashok:2020dnc}
S.~K.~Ashok and J.~Troost,
``Superstrings in Thermal Anti-de Sitter Space,''
[arXiv:2012.08404 [hep-th]].
%0 citations counted in INSPIRE as of 12 Jan 2021







%\cite{Giveon:2003ku}
\bibitem{Giveon:2003ku}
A.~Giveon and A.~Pakman,
``More on superstrings in AdS(3) x N,''
JHEP \textbf{03} (2003), 056
doi:10.1088/1126-6708/2003/03/056
[arXiv:hep-th/0302217 [hep-th]].
%31 citations counted in INSPIRE as of 27 Jan 2021

%\cite{Gaberdiel:2017oqg}
\bibitem{Gaberdiel:2017oqg}
M.~R.~Gaberdiel, R.~Gopakumar and C.~Hull,
``Stringy AdS$_{3}$ from the worldsheet,''
JHEP \textbf{07} (2017), 090
doi:10.1007/JHEP07(2017)090
[arXiv:1704.08665 [hep-th]].
%33 citations counted in INSPIRE as of 08 Feb 2021


\bibitem{FarkasKra}
H.~Farkas, I.~Kra, ``Theta constants, Riemann surfaces and the modular group,''
Graduate studies in mathematics, Vol.37. Amer. Math. Soc. 2001.


%\cite{Maldacena:2000hw}
\bibitem{Maldacena:2000hw}
J.~M.~Maldacena and H.~Ooguri,
``Strings in AdS(3) and SL(2,R) WZW model 1.: The Spectrum,''
J. Math. Phys. \textbf{42} (2001), 2929-2960
doi:10.1063/1.1377273
[arXiv:hep-th/0001053 [hep-th]].
%474 citations counted in INSPIRE as of 04 Feb 2021

%\cite{Seiberg:1999xz}
\bibitem{Seiberg:1999xz}
N.~Seiberg and E.~Witten,
``The D1 / D5 system and singular CFT,''
JHEP \textbf{04} (1999), 017
doi:10.1088/1126-6708/1999/04/017
[arXiv:hep-th/9903224 [hep-th]].
%479 citations counted in INSPIRE as of 29 Jan 2021



%\cite{Krasnov:2000zq}\cite{Li:2019mwb}
\bibitem{Krasnov:2000zq}
K.~Krasnov,
``Holography and Riemann surfaces,''
Adv. Theor. Math. Phys. \textbf{4} (2000), 929-979
doi:10.4310/ATMP.2000.v4.n4.a5
[arXiv:hep-th/0005106 [hep-th]].
%113 citations counted in INSPIRE as of 27 Jan 2021

%\cite{Li:2019mwb}
\bibitem{Li:2019mwb}
S.~Li, N.~Toumbas and J.~Troost,
``Liouville Quantum Gravity,''
Nucl. Phys. B \textbf{952} (2020), 114913
doi:10.1016/j.nuclphysb.2019.114913
[arXiv:1903.06501 [hep-th]].
%4 citations counted in INSPIRE as of 27 Jan 2021

%\cite{Argurio:2000tb}
\bibitem{Argurio:2000tb}
R.~Argurio, A.~Giveon and A.~Shomer,
``Superstrings on AdS(3) and symmetric products,''
JHEP \textbf{12} (2000), 003
doi:10.1088/1126-6708/2000/12/003
[arXiv:hep-th/0009242 [hep-th]].
%97 citations counted in INSPIRE as of 23 Feb 2021

\bibitem{Tseytlin:1988rr}
A.~A.~Tseytlin,
``Sigma Model Approach to String Theory,''
Int. J. Mod. Phys. A \textbf{4} (1989), 1257
doi:10.1142/S0217751X8900056X.
%130 citations counted in INSPIRE as of 11 Feb 2021 




%\cite{Giveon:1998ns}\cite{Kutasov:1999xu}
\bibitem{Giveon:1998ns}
A.~Giveon, D.~Kutasov and N.~Seiberg,
``Comments on string theory on AdS(3),''
Adv. Theor. Math. Phys. \textbf{2} (1998), 733-782
doi:10.4310/ATMP.1998.v2.n4.a3
[arXiv:hep-th/9806194 [hep-th]].
%394 citations counted in INSPIRE as of 04 Aug 2020
%\cite{Kutasov:1999xu}
\bibitem{Kutasov:1999xu}
D.~Kutasov and N.~Seiberg,
``More comments on string theory on AdS(3),''
JHEP \textbf{04} (1999), 008
doi:10.1088/1126-6708/1999/04/008
[arXiv:hep-th/9903219 [hep-th]].
%219 citations counted in INSPIRE as of 04 Aug 2020


%\cite{Li:2020nei}
\bibitem{Li:2020nei}
S.~Li and J.~Troost,
``Twisted String Theory in Anti-de Sitter Space,''
JHEP \textbf{11} (2020), 047
doi:10.1007/JHEP11(2020)047
[arXiv:2005.13817 [hep-th]].
%4 citations counted in INSPIRE as of 26 Nov 2020







%\cite{Troost:2011ud}
\bibitem{Troost:2011ud}
J.~Troost,
``The $AdS_3$ central charge in string theory,''
Phys. Lett. B \textbf{705} (2011), 260-263
doi:10.1016/j.physletb.2011.10.007
[arXiv:1109.1923 [hep-th]].
%7 citations counted in INSPIRE as of 08 Jan 2021

% %\cite{Brown:1986nw}
\bibitem{Brown:1986nw}
J.~D.~Brown and M.~Henneaux,
``Central Charges in the Canonical Realization of Asymptotic Symmetries: An Example from Three-Dimensional Gravity,''
Commun. Math. Phys. \textbf{104} (1986), 207-226
doi:10.1007/BF01211590.
% %1885 citations counted in INSPIRE as of 02 Nov 2020

%\cite{Maldacena:2001km}
\bibitem{Maldacena:2001km}
J.~M.~Maldacena and H.~Ooguri,
``Strings in AdS(3) and the SL(2,R) WZW model. Part 3. Correlation functions,''
Phys. Rev. D \textbf{65} (2002), 106006
doi:10.1103/PhysRevD.65.106006
[arXiv:hep-th/0111180 [hep-th]].
%241 citations counted in INSPIRE as of 01 Apr 2021

%\cite{Buchbinder:2011jr}
\bibitem{Buchbinder:2011jr}
E.~I.~Buchbinder and A.~A.~Tseytlin,
``Semiclassical correlators of three states with large $S^5$ charges in string theory in $AdS_5 \times S^5$,''
Phys. Rev. D \textbf{85} (2012), 026001
doi:10.1103/PhysRevD.85.026001
[arXiv:1110.5621 [hep-th]].
%39 citations counted in INSPIRE as of 01 Apr 2021



\end{thebibliography}

\end{document}